\documentclass[reprint, two column, prl, aps, showkeys]{revtex4-1}

\usepackage{amssymb}
\usepackage{amsmath}
\usepackage{graphicx}
\usepackage{dcolumn}
\usepackage{bm}
\usepackage{color}
\usepackage{ulem}
\usepackage{float}
\usepackage{textcomp}
\restylefloat{table}
\usepackage{verbatim}
\usepackage{epstopdf}
\usepackage[mediumspace,mediumqspace,squaren]{SIunits}

\begin{document}

\title{Direct observation of composite fermions and their fully spin-polarized Fermi sea near $\nu=5/2$}
\date{\today}

\author{Md.\ Shafayat Hossain}
\author{Meng K.\ Ma}
\author{M. A.\ Mueed}
\author{L. N.\ Pfeiffer} 
\author{K. W.\ West}
\author{K. W.\ Baldwin}
\author{M.\ Shayegan}
\affiliation{Department of Electrical Engineering, Princeton University, Princeton, New Jersey 08544, USA}

\begin{abstract}
The enigmatic even-denominator fractional quantum Hall state at Landau level filling factor $\nu=5/2$ is arguably the most promising candidate for harboring Majorana quasi-particles with non-Abelian statistics and thus of potential use for topological quantum computing. The theoretical description of the  $\nu=5/2$ state is generally believed to involve a topological $p$-wave pairing of fully spin-polarized composite fermions through their condensation into a non-Abelian Moore-Read Pfaffian state. There is, however, no direct and conclusive experimental evidence for the existence of composite fermions near $\nu=5/2$ or for an underlying fully spin-polarized Fermi sea.  Here, we report the observation of composite fermions very near $\nu=5/2$ through geometric resonance measurements, and find that the measured Fermi wavevector provides direct demonstration of a Fermi sea with full spin polarization. This lends crucial credence to the model of $5/2$ fractional quantum Hall effect as a topological $p$-wave paired state of composite fermions. 
\end{abstract} 

\maketitle

Since the discovery of the fractional quantum Hall effect (FQHE) \cite{Tsui.PRL.1982}, a low-disorder two-dimensional electron system (2DES) has been the platform of choice to investigate the interplay between quantum mechanics and electron-electron interaction \cite{Jain.2007}. The FQHE states typically form at high perpendicular magnetic fields ($B$) in the lowest ($N = 0$) Landau level (LL)  when the filling factor ($\nu$) assumes fractional values with odd denominators \cite{Jain.2007}. These FQHE states can be elegantly explained in the composite fermion (CF) theory \cite{Jain.PRL.1989, Jain.2007} where an electron merges with an even number of flux quanta to form an exotic quasi-particle which experiences a zero \textit{effective} magnetic field ($B^{*} = 0$) exactly at the $1/2$ filling. The FQHE is then explained as the integer QHE of CFs. Moreover, the CFs near $\nu=1/2$ occupy a Fermi sea and can execute cyclotron motion at small $B^{*}$, similar to their fermion counterparts near $B=0$ \cite{Halperin.PRB.1993}. With the application of a periodic potential or density perturbation, if the CFs can complete a cyclotron orbit without scattering, then they exhibit a geometric resonance (GR) when their orbit diameter equals the period of the perturbation. Indeed, the observation of such a resonance has provided the most direct and compelling evidence for the presence of CFs near $\nu=1/2$ \cite{Willett.PRL.1993, Kang.PRL.1993, Smet.PRL.1999, Kamburov.PRB.2014}. 

In contrast to the FQHE states in the lowest LL, the FQHE observed in the first-excited ($N=1$) LL at the \textit{even-denominator} filling $\nu=5/2$ has remained enigmatic since its discovery \cite{Willett.PRL.1987}. The initial explanation \cite{Willett.PRL.1987} for the $5/2$ FQHE state was based on a two-component, spin-unpolarized, $\Psi_{331}$ wavefunction \cite{Halperin.Helv.Phys.Acta.1983}. Following this argument, Haldane and Rezayi \cite{Haldane.PRL.1988} constructed a spin-unpolarized, CF wavefunction based on the hollow-core model. On the other hand, Moore and Read \cite{Moore.Nucl.Phys.B.1991} presented a spin-polarized, $p$-wave-paired, CF ground state, known as the Moore-Read Pfaffian. Spectacularly, this state should harbor Majorana quasi-particles obeying non-Abelian statistics whose interchange takes the system from one of its many ground states to another, whereas the interchange of ordinary Abelian quasi-particles only adds a phase to their wavefunctions. Numerical investigations \cite{ Morf.PRL.1998} support such a non-Abelian ground state at $\nu=5/2$. As a consequence, the $5/2$ FQHE has attracted much attention as a promising platform for topological quantum computation \cite{Nayak.Rev.Mod.Phys.2008}.

\begin{figure}[t!]
\includegraphics[width=0.47\textwidth]{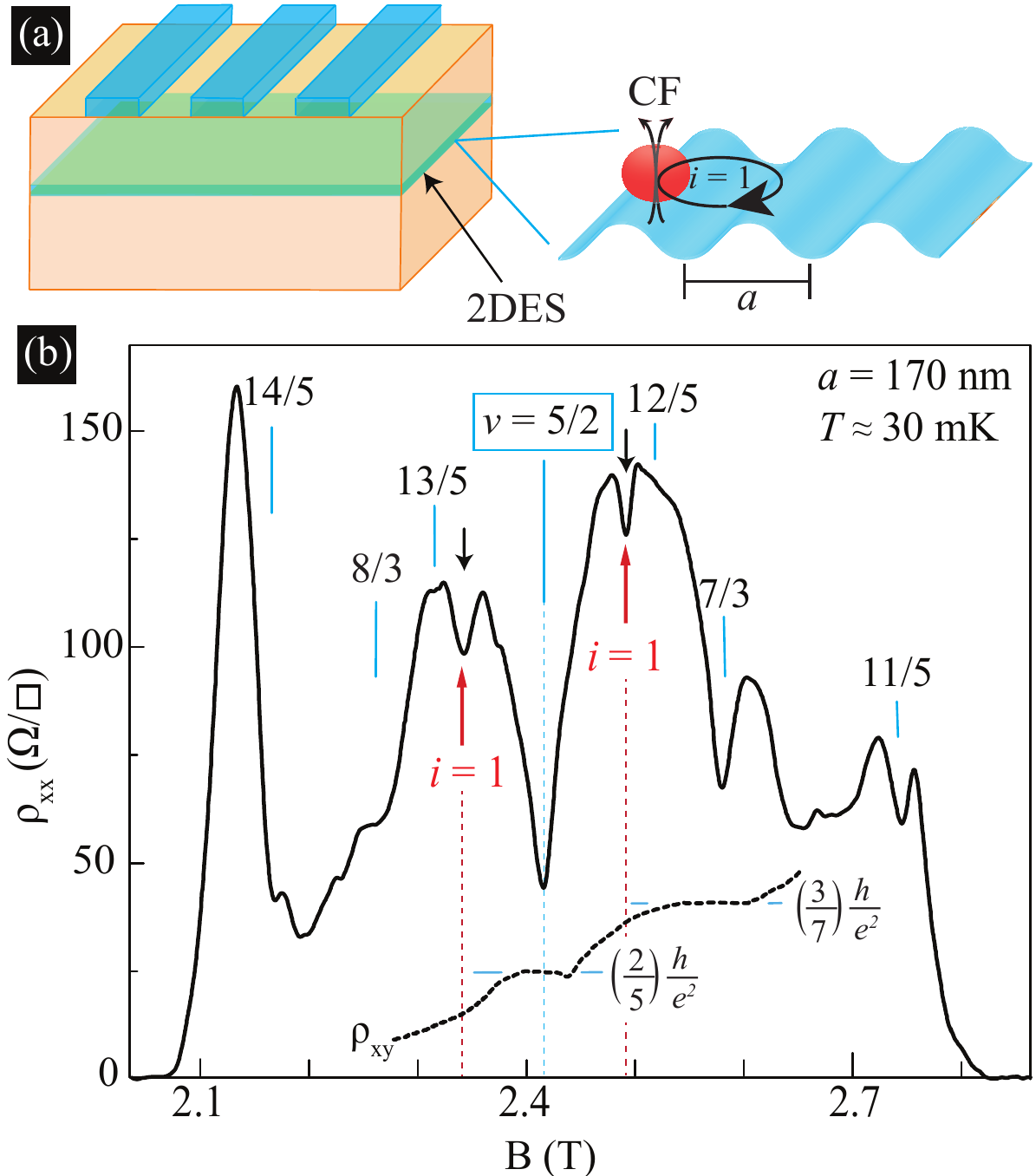}
\caption{\label{fig:Fig1} GR features for CFs near $\nu=5/2$. (a) Lateral surface superlattice, inducing a periodic density perturbation in the 2DES. When the cyclotron orbit of the CFs become commensurate with the period of the perturbation, the $i = 1$ GR occurs (see text). (b) Magnetoresistance trace for the first-excited LL revealing the $i = 1$ CF GR features, resistance minima marked with black arrows flanking $\nu=5/2$. These minima do not coincide with any of the expected FQHE minima which are labeled by blue markers. Blue horizontal lines mark the expected quantized values of the Hall plateau for $\nu=5/2$ and $7/3$ in the $\rho_{xy}$ trace. Red arrows indicate the \textit{expected} positions for the $i = 1$ GR for fully spin-polarized CFs near $\nu=5/2$.
}
\end{figure}  

\begin{figure}[t!]
\includegraphics[width=.47\textwidth]{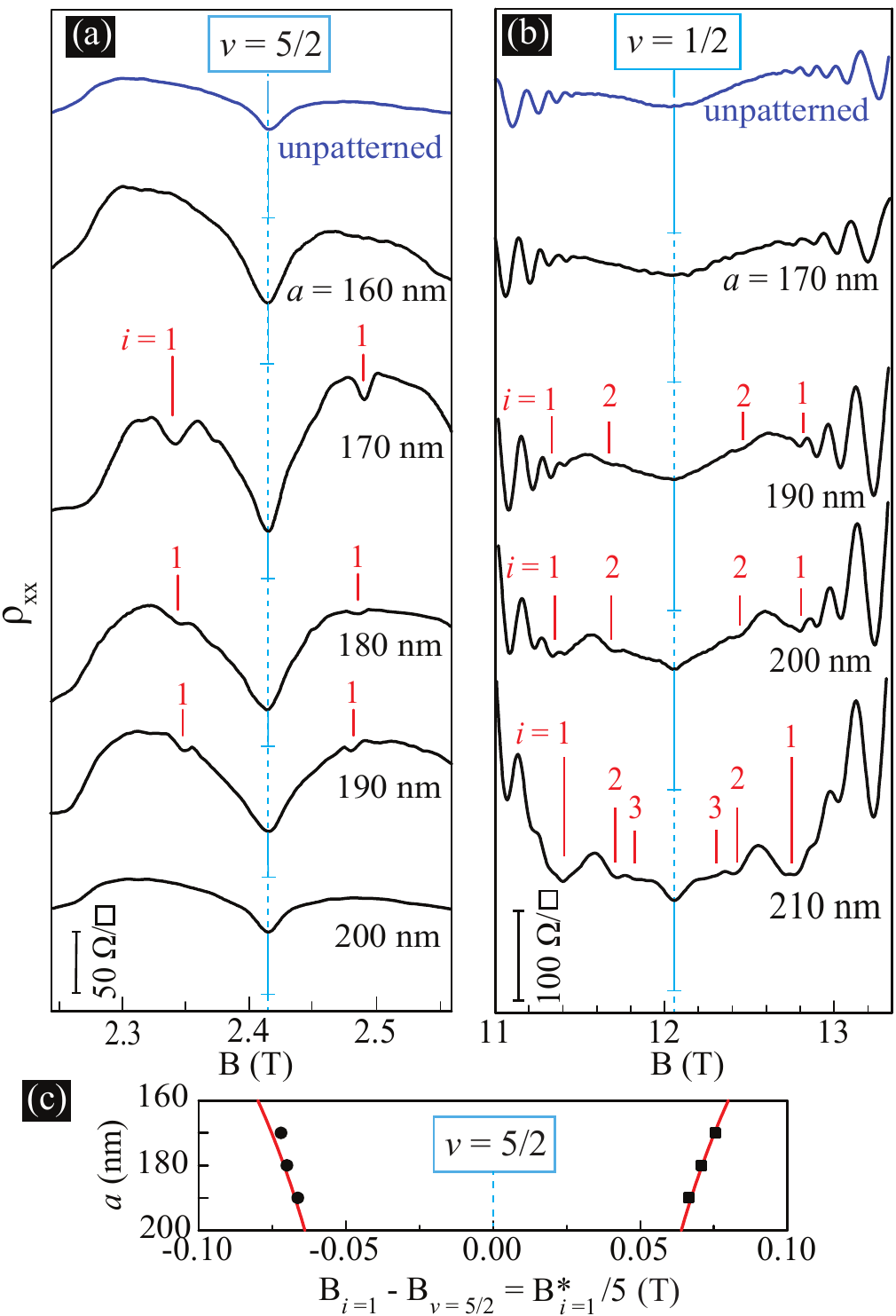}
\caption{\label{fig:Fig2} Dependence of the GR features of CFs near (a) $\nu = 5/2$ and (b) $\nu = 1/2$ on the period of the surface superlattice. Magnetoresistance traces are vertically offset for clarity and the zero-resistance positions are indicated with blue horizontal marks. The $expected$ positions for GRs are marked with red lines. (c) Summary of the positions of the observed GR minima (symbols) and expected values (lines) near $\nu = 5/2$.}
\end{figure}

Despite the enormous interest in the $5/2$ state, little conclusive experimental evidence is available for its origin \cite{Willett.RPP.2013}. Early surface acoustic wave experiments \cite{Willett.PRL.2002,Willett.RPP.2013} showed hints of Fermi surface effects at $\nu=5/2$, but there has been no compelling evidence for the existence of CFs near $\nu=5/2$. Based on the quasi-particle tunneling and interference studies \cite{Radu.Science.2008, Lin.PRB.2012, Willett.RPP.2013}, it is still unclear whether the ground state at $\nu =5/2$ is Abelian or non-Abelian. The most important prerequisite for the $5/2$ FQHE to be non-Abelian is that it is single-component (fully spin-polarized). However, the spin polarization of the $5/2$ state has remained an open question \cite{Willett.RPP.2013} even 30 years after its discovery. Initial tilted-field measurements \cite{Eisenstein.PRL.1988} indicated the $\nu = 5/2$ FQHE to be spin unpolarized, a conclusion which is also favored by the photoluminescence \cite{Stern.PRL.2010} and inelastic light scattering \cite {Wurstbauer.2012} experiments. On the other hand, density-dependence \cite{Pan.SSC.2001}, and later tilt-dependence \cite{Zhang.PRL.2010} activation gap measurements suggested possible full spin polarization. This observation was supported in recent nuclear magnetic resonance \cite{Tiemann.Science.2012, Stern.PRL.2012} and tunneling experiments \cite{Eisenstein.PRL.2017}. 

Here we address two fundamental questions regarding the origin of the $5/2$ FQHE state: Are CFs present near $\nu=5/2$, and are they fully spin-polarized? Our GR measurements, which directly probe the Fermi sea and wavevector of CFs, provide unambiguous and conclusive positive answers to these questions. We emphasize that our technique is simple, and yet most direct and quantitative; also, it does not rely on any fitting schemes or parameters. Moreover, our measurements do not involve exposure of the sample to illumination/radiation, or to in-plane magnetic fields.

Our experimental platform is a clean 2DES with density $1.46\times10^{11}$ cm$^{-2}$ confined to a modulation-doped, 30-nm-wide, GaAs quantum well; see Supplementary Materials for more details of the sample structure. In our GR measurements, we employ a weak perturbation in the form of a minute periodic density modulation, the estimated magnitude of which is about $<<1\%$ \cite{Mueed.PRL.2016}. As illustrated in Fig. 1(a), this is achieved by fabricating a one-dimensional array of stripes of negative electron-beam resist on the surface of the lithographically defined Hall bar \cite{Kamburov.PRB.2014, Mueed.PRL.2016, Davies.PRB.1994, Kamburov3/2.PRB.2014, Kamburov.PRL.2014, Mueed.PRB.2017}. Thanks to the piezoelectric effect, the strain from this surface superlattice propagates to the 2DES and leads to a minute density perturbation. 

\begin{figure*}[t!]
\includegraphics[width=.98\textwidth]{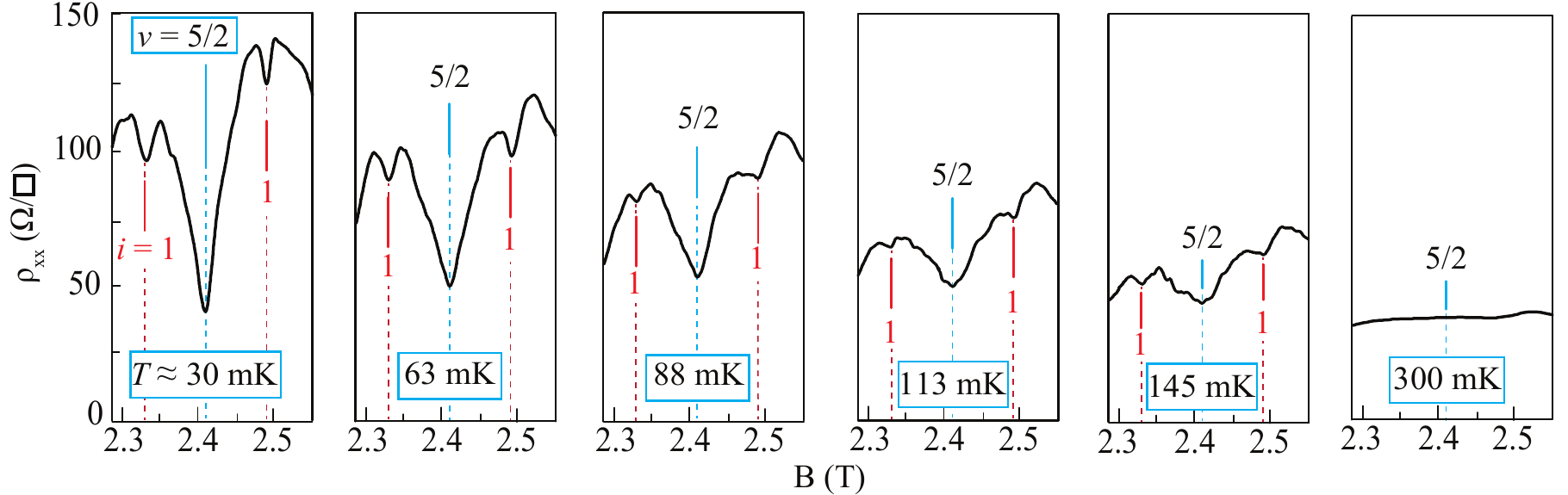}
\caption{\label{fig:Fig3} Evolution of the GR of CFs near $\nu=5/2$ with temperature. Clear GR minima are observed at 30 mK, accompanied by a strong FQHE state at $\nu = 5/2$. As the temperature is raised, the GR minima, together with the V-shaped minimum at $\nu = 5/2$ progressively become weaker and disappear completely at 300 mK.}
\end{figure*}

Weakly-interacting CFs subjected to an effective magnetic field $B^{*}$ execute cyclotron motion with an orbit diameter of $2R_{c}^{*} = 2\hbar k_F^{*}/eB^{*}$, the size of which is determined by the Fermi wavevector of the CFs, $k_F^{*}$. For a fully spin-polarized CF Fermi sea, $k_F^{*}=(4\pi n_{CF}^{*})^{1/2}$ where $n_{CF}^{*}$ is the CF density, whereas for a spin-unpolarized CF Fermi sea, $k_F^{*}=(2\pi n_{CF}^{*})^{1/2}$. If the CFs have a sufficiently long mean-free-path so they can complete a cyclotron orbit without scattering, then a GR occurs when the orbit diameter matches the period ($a$) of the perturbation; see Fig. 1(a) for a schematic illustration. More quantitatively \cite{Smet.PRL.1999, Kamburov.PRB.2014, Kamburov3/2.PRB.2014, Kamburov.PRL.2014, Mueed.PRB.2017}, when $2R_{c}^{*}/a=i+1/4$ ($i=1,2,3,...$), GRs manifest as minima in magnetoresistance at $B_{i}^{*}=2\hbar k_F^{*}/ea(i+1/4)$. Thus, $k_F^{*}$ can be deduced directly from the positions of $B_{i}^{*}$. Observation of such well-defined GR conclusively proves the presence of the CF Fermi sea and also provides a direct measure of the CF spin-polarization. This is well established for CFs near the compressible states at $\nu=1/2$ and $3/2$ \cite{Smet.PRL.1999, Kamburov.PRB.2014, Kamburov3/2.PRB.2014, Kamburov.PRL.2014, Mueed.PRB.2017}. Here, we apply the GR technique to the $\nu=5/2$ FQHE state.

The most important finding of our work is highlighted in Fig. 1(b). This magnetotransport trace for the first-excited LL exhibits well-developed GR minima (marked with two black arrows) flanking the deep V-shaped minimum at $\nu=5/2$. The presence of these GR features confirms the existence of a well-defined Fermi wavevector, thus providing direct evidence for a CF Fermi sea. 

From the period of the perturbation, $a = 170$ nm, and assuming that the magnetoresistance minima we observe stem from the primary $i = 1$ GR, we can determine $k_F^{*}$ according to $k_F^{*} = B_{i=1}^{*} ea(1+1/4)/2\hbar$ where $B_{i=1}^*=5(B_{i=1}-B_{\nu=5/2})$; $B_{i=1}$ is the magnetic field at which we observe the resistance minimum. We obtain $k_F^{*}$ = $6.12 \times 10^{5}$ cm$^{-1}$ from the right minimum and $k_F^{*}$ = $5.99 \times 10^{5}$ cm$^{-1}$ from the left minimum. These values are in excellent agreement with $k_F^{*}=(4\pi n_{CF}^{*})^{1/2}$ = $6.06 \times 10^{5}$ cm$^{-1}$, expected for fully spin-polarized CFs; note that the CF density near $\nu =5/2$ is $n_{CF}^{*} = n_e/5$, where $n_e =1.46\times10^{11}$ cm$^{-2}$ is the 2DES electron density. On the other hand, spin-unpolarized CFs are expected to have a $k_F^{*}=(2\pi n_{CF}^{*})^{1/2}$ = $4.27 \times 10^{5}$ cm$^{-1}$, clearly different from our observation. The data therefore provide direct and conclusive evidence for CFs near $\nu=5/2$ and their full spin-polarization. We have marked in Fig. 1(b) the \textit{expected} $i = 1$ GR positions corresponding to fully spin-polarized CFs with red arrows.

It is clear from Fig. 1(b) that the positions of the GR features do not coincide with any of the observed or expected odd-denominator FQHE minima (marked with blue lines). Furthermore, the FQHE at $\nu=5/2$ is well developed as is evident from the strong minimum in $\rho_{xx}$ and the corresponding plateau at $(2/5)(h/e^2)$ in the Hall ($\rho_{xy}$) trace. The coexistence of the FQHE state at $\nu=5/2$ and fully spin-polarized CFs on its sides strongly argues in favor of a fully spin-polarized $\nu=5/2$ FQHE state.

In Fig. 2(a) we show the measured magnetoresistance traces for six different sections of the Hall bar: an unpatterned (top trace), and five patterned sections with surface superlattice periods $a$ ranging from 160 to 200 nm \cite{footnote1}. The $i =1$ GR minima, marked by vertical red lines, are observed only in a narrow range of $a$. They are strongest for $a=170$ nm. The $a=180$ and $190$ nm sections exhibit very weak GR minima whose positions are consistent with the expected values (Fig. 2(c)), whereas no GR features appear for $a=160$ or $\geq200$ nm (traces for $a\geq$ 200 nm are not shown). A simple interpretation of the disappearance of minima for $a\geq$ 200 nm is that the CFs near $\nu=5/2$ are very fragile and have a very short mean-free-path, so that they can complete (ballistically) only cyclotron orbits which are small. The absence of minima for the shortest period $a=160$ nm can be attributed to the ever decreasing amplitude of the periodic density modulation as its period becomes small compared to the depth of the 2DES below the surface (235 nm for our sample) \cite{Davies.PRB.1994, Mueed.PRL.2016}.

We present in Fig. 2(b) traces which were taken in the same sample but near filling $\nu = 1/2$. As expected, and seen in the top trace (unpatterned region) of Fig. 2(b), at $\nu=1/2$ the 2DES is compressible and shows a smooth, broad minimum. It hosts a sea of CFs with density $n_{CF}^{*} = n_e$ and, as revealed in the lowest trace of Fig. 2(b), exhibits strong GR minima at $i=1, 2$ and $3$ for $a=210$ nm. The sample also exhibits clear GR for $a=200$ nm, but for smaller periods the GR minima become weaker and are essentially absent for the $a=170$ nm trace in Fig. 2(b). The contrast between the traces near $\nu=5/2$ and $1/2$ for $a=170$ nm in Fig. \ref{fig:Fig2} is remarkable. It suggests that the CFs near $\nu=5/2$ require a very gentle potential modulation, gentler than the $\nu=1/2$ CFs, to exhibit GR features.

Two noteworthy differences between the CFs near $\nu=5/2$ and $1/2$ are their density and size. Near $\nu=1/2$, the CF density is nearly equal to the electron density \cite{Kamburov.PRL.2014}, while near $\nu=5/2$ it is only $1/5$ of the electron density (see Supplementary Materials for a discussion of CF density). This much smaller density could lead to a shorter mean-free-path for the CFs near $5/2$. A recent GR study \cite{Mueed.PRB.2017}, however, indicated that it is not the smaller CF density but rather the higher LL index that plays the dominant role. Indeed, Ref. \cite{Mueed.PRB.2017} showed that, in two-subband 2DESs 
confined to wide GaAs quantum wells, where $\nu=5/2$ occurs in the $N=0$ LL (of the anti-symmetric subband), the CFs show clear GRs. In contrast, in similar quality but narrower quantum well samples, where $\nu=5/2$ was in the $N=1$ LL, no GR features were observed (for $T\geq$ 300 mK). We note that, in the $N=1$ LL, the quasi-particles for the FQHE states near $\nu=5/2$ (e.g., at $\nu=7/3$) are theoretically expected to be much larger than their counterparts in the $N=0$ LL near $\nu=1/2$ (e.g., at $\nu=1/3$) \cite{Balram.PRL.2013, Johri.PRB.2014}. If the larger size also applies to the CFs, it could be another reason for the extreme fragility of the $\nu=5/2$ CFs and why the observation of their GR had remained elusive until now.

Figure 3 captures the temperature dependence of the GR minima near $\nu = 5/2$. The minima are best pronounced at the lowest temperature, slowly become weaker at higher temperatures, and completely disappear at 300 mK; see Fig. S4 in Supplementary Materials for larger field-range and comparison with the unpatterned section. Note that at 300 mK, the FQHE minimum at $\nu=5/2$ also disappears. It is noteworthy again, that in the case of the $\nu=1/2$ (and $\nu=3/2$) CFs in the $N=0$ LL, their GR minima are very strong even at $300$ mK \cite{Kamburov.PRB.2014, Kamburov3/2.PRB.2014, Kamburov.PRL.2014, Mueed.PRB.2017}, further attesting to their robustness compared to the $\nu=5/2$ CFs.

It is tempting to interpret the weakening of the GR features observed in Fig. 3 as a signal that the $\nu=5/2$ CFs are gradually becoming spin-unpolarized at higher temperatures, as hinted at in nuclear magnetic resonance experiments \cite{Tiemann.Science.2012}. Partially spin-polarized CFs, however, should populate two Fermi seas with smaller Fermi wavevectors. If so, as reported for partially spin-polarized CFs near $\nu=3/2$ \cite{Kamburov3/2.PRB.2014}, we would expect the GR minima to become broad and also move to smaller values of $B^{*}$ (closer to $\nu=5/2$). This is in contrast to the fixed positions of the GR minima in Fig. 3 as temperature is raised. It is more likely that the weakening we observe at high temperatures is a reflection of the much smaller Fermi energy of the CFs near $\nu=5/2$ compared to the $\nu=1/2$ CFs. Note that, qualitatively, the Fermi energy for CFs should scale with the Coulomb energy, $e^2/l_B$, where $l_B=\sqrt{\hbar/eB}$, implying a factor of $\sim\sqrt5$ smaller Fermi energy for the $\nu=5/2$ CFs \cite{Jain.2007}.

Our direct measurements of the CF Fermi wavevector provide compelling evidence for the existence of a fully spin-polarized Fermi sea of CFs near $\nu=5/2$. Insofar as the non-Abelian state theoretically expected at $\nu=5/2$ entails a pairing of fully spin-polarized CFs and their condensation into a FQHE state, our data offer a significant milestone since they demonstrate the presence of CFs and their full spin polarization very near $\nu=5/2$. Our findings should foster further efforts to experimentally demonstrate the non-Abelian nature of the $\nu=5/2$ quasi-particles, and implement their braiding for fault-tolerant topological quantum computation.

\begin{acknowledgments}
We acknowledge support through the NSF (Grants DMR 1709076 and ECCS 1508925) for measurements, and the NSF (Grant MRSEC DMR 1420541), the DOE BES (Grant DE-FG02-00-ER45841), and the Gordon and Betty Moore Foundation (Grant GBMF4420 for sample fabrication and characterization. This research is funded in part by QuantEmX grants from ICAM and the Gordon and Betty Moore Foundation through Grant GBMF5305 to M. S. H., M. K. M., and M. S. Our measurements were partly performed at the National High Magnetic Field Laboratory (NHMFL), which is supported by the NSF Cooperative Agreement DMR 1157490, by the State of Florida, and the DOE. We thank S. Hannahs, T. Murphy, H. Baek, J. Park and G. Jones at NHMFL for technical support. We also thank J. K. Jain for illuminating discussions.
\end{acknowledgments}


\begin{thebibliography}{99}

\bibitem {Tsui.PRL.1982} D. C. Tsui, H. L. Stormer, and A. C. Gossard, \textit{Phys. Rev. Lett.} \textbf{48}, 1559 (1982).

\bibitem {Jain.2007}  J. K. Jain, \textit{Composite fermions}.
(Cambridge University Press, New York, 2007).

\bibitem {Jain.PRL.1989}  J. K. Jain, \textit{Phys. Rev. Lett.} \textbf{63}, 199 (1989).

\bibitem{Halperin.PRB.1993} B. I. Halperin, P. A. Lee, and N. Read, \textit{Phys. Rev. B} \textbf{47}, 7312 (1993).

\bibitem{Willett.PRL.1993} R. L. Willett, R. R. Ruel, K. W. West, and L. N. Pfeiffer, \textit{Phys. Rev. Lett.} \textbf{71}, 3846 (1993).

\bibitem{Kang.PRL.1993} W. Kang, H. L. Stormer, L. N. Pfeiffer, K. W. Baldwin, and K. W. West, \textit{Phys. Rev. Lett.} \textbf{71}, 3850 (1993).

\bibitem {Smet.PRL.1999}  J. H. Smet, S. Jobst, K. von Klitzing, D. Weiss, W. Wegscheider, and V. Umansky, \textit{Phys. Rev. Lett.} \textbf{83}, 2620 (1999).

\bibitem {Kamburov.PRB.2014} D. Kamburov, M. A. Mueed, M. Shayegan, L. N. Pfeiffer, K. W. West, K. W. Baldwin, J. J. D. Lee, and R. Winkler, \textit{Phys. Rev. B} \textbf{89}, 085304 (2014).

\bibitem {Willett.PRL.1987} R. L. Willett, J. P. Eisenstein, H. L. Stormer, D. C. Tsui, A. C. Gossard, and J. H. English,  \textit{Phys. Rev. Lett.} \textbf{59}, 1776 (1987).

\bibitem {Halperin.Helv.Phys.Acta.1983} B. I. Halperin, \textit{Helv. Phys. Acta} \textbf{56}, 75 (1983).

\bibitem {Haldane.PRL.1988} F. D. M. Haldane and E. H. Rezayi, \textit{Phys. Rev. Lett.} \textbf{60}, 956 (1988).

\bibitem {Moore.Nucl.Phys.B.1991} G. Moore, and N. Read, \textit{Nucl. Phys. B} \textbf{360}, 362 (1991).

\bibitem {Morf.PRL.1998}  R. Morf, \textit{Phys. Rev. Lett.} \textbf{80}, 1505 (1998).

\bibitem {Nayak.Rev.Mod.Phys.2008} C. Nayak, S. H. Simon, A. Stern, M. Freedman, and S. Das Sarma, \textit{Rev. Mod. Phys.} \textbf{80}, 1083 (2008).

\bibitem{Willett.RPP.2013}R. L. Willett, \textit{Rep. Prog. Phys.} \textbf{76}, 076501 (2013).

\bibitem {Willett.PRL.2002}  R. L. Willett, K. W. West, and L. N. Pfeiffer, \textit{Phys. Rev. Lett.} \textbf{88}, 066801 (2002).

\bibitem {Radu.Science.2008}  I. P. Radu, J. B. Miller, C. M. Marcus, M. A. Kastner, L. N. Pfeiffer, and K. W. West, \textit{Science} \textbf{320}, 899 (2008).

\bibitem {Lin.PRB.2012} X. Lin, C. Dillard, M. A. Kastner, L. N. Pfeiffer, and K. W. West, \textit{Phys. Rev. B} \textbf{85}, 165321 (2012).

\bibitem {Eisenstein.PRL.1988}J. P. Eisenstein, R. Willett, H. L. Stormer, D. C. Tsui, A. C. Gossard, and J. H. English, \textit{Phys. Rev. Lett.} \textbf{61}, 997 (1988).

\bibitem {Stern.PRL.2010}M. Stern, P. Plochocka, V. Umansky, D. K. Maude, M. Potemski and I. Bar-Joseph,  \textit{Phys. Rev. Lett.} \textbf{105} 096801 (2010).

\bibitem {Wurstbauer.2012}U. Wurstbauer, K. W. West, L. N. Pfeiffer, and A. Pinczuk, \textit{Phys. Rev. Lett.} \textbf{110}, 026801 (2013).

\bibitem {Pan.SSC.2001} W. Pan, H. L. Stormer, D. C. Tsui, L. N. Pfeiffer, K. W. Baldwin and K. W. West,  \textit{Solid State Commun.} \textbf{119} 641 (2001).

\bibitem{Zhang.PRL.2010} Chi Zhang, T. Knuuttila, Yanhua Dai, R. R. Du, L. N. Pfeiffer, and K. W. West, \textit{Phys. Rev. Lett.} \textbf{104}, 166801 (2010).

\bibitem {Tiemann.Science.2012} L. Tiemann, G.	Gamez, N.	Kumada, and K.	Muraki, \textit{Science} \textbf{335}, 6070 (2012).

\bibitem {Stern.PRL.2012} M. Stern, B. A. Piot, Y. Vardi, V. Umansky, P. Plochocka, D. K. Maude and I. Bar-Joseph, \textit{Phys. Rev. Lett.} \textbf{108} 066810 (2012).

\bibitem {Eisenstein.PRL.2017}  J. P. Eisenstein, L. N. Pfeiffer, and K. W. West, \textit{Phys. Rev. Lett.} \textbf{118}, 186801 (2017).

\bibitem {Mueed.PRL.2016} M. A. Mueed, Md. Shafayat Hossain, L. N. Pfeiffer, K. W. West, K. W. Baldwin, and M. Shayegan, \textit{Phys. Rev. Lett.} \textbf{117}, 076803 (2016).
 
\bibitem{Davies.PRB.1994} J. H. Davies and I. A. Larkin, \textit{Phys. Rev. B} \textbf{49}, 4800 (1994).

\bibitem{Kamburov3/2.PRB.2014} D. Kamburov, M. A. Mueed, I. Jo, Yang Liu, M. Shayegan, L. N. Pfeiffer, K. W. West, K. W. Baldwin, J. J. D. Lee, and R. Winkler, \textit{Phys. Rev. B} \textbf{90}, 235108 (2014).

\bibitem {Kamburov.PRL.2014} D. Kamburov, Yang Liu, M. A. Mueed, M. Shayegan, L. N. Pfeiffer, K. W. West, and K. W. Baldwin, \textit{Phys. Rev. Lett.} \textbf{113}, 196801 (2014).

\bibitem {Mueed.PRB.2017} M. A. Mueed, D. Kamburov, Md. Shafayat Hossain, L. N. Pfeiffer, K. W. West, K. W. Baldwin, and M. Shayegan, \textit{Phys. Rev. B} \textbf{95}, 165438 (2017).

\bibitem {footnote1} In Fig. S3 of Supplementary Materials, we show the same data in a larger field range, along with the corresponding Hall traces.
 

\bibitem {Balram.PRL.2013} Ajit C. Balram, Ying-Hai Wu, G. J. Sreejith, Arkadiusz W\'ojs, and Jainendra K. Jain, \textit{Phys. Rev. Lett.} \textbf{110}, 186801 (2013).


\bibitem {Johri.PRB.2014} Sonika Johri, Z. Papi\'c, R. N. Bhatt, and P. Schmitteckert, \textit{Phys. Rev. B} \textbf{89}, 115124 (2014).


\end{thebibliography}
\end{document}